\documentclass[runningheads]{llncs}
\usepackage[T1]{fontenc}
\usepackage{graphicx}
\usepackage{xspace}
\usepackage{amsmath}
\usepackage{gensymb} % \degree
\usepackage{microtype}
\usepackage{cite}
\newcommand{\parag}[1]{\noindent\textbf{#1}\quad}
\newcommand{\methodit}[1]{\textit{\underline{#1}}}

\newcommand{\tagnet}{\ensuremath{\theta_{\mathrm{tag}}}\xspace}
\newcommand{\pinnt}{\ensuremath{\theta_{\mathrm{pinn},t}}\xspace}
\newcommand{\pinntpo}{\ensuremath{\theta_{\mathrm{pinn},t+1}}\xspace}
\newcommand{\hvset}{\ensuremath{\{h,v\}}\xspace}

\DeclareMathOperator*{\argmin}{arg\,min\xspace}

\begin{document}
\title{Brightness-Invariant Tracking Estimation \\in Tagged MRI}

\titlerunning{BRITE: Brightness-Invariant Tracking Estimation}

% \orcidID{0000-1111-2222-3333}

\author{Zhangxing Bian\inst{1}\thanks{corresponding author: zbian4@jhu.edu} \and Shuwen Wei\inst{1} \and  Xiao Liang\inst{2} \and Yuan-Chiao Lu\inst{3,4} \and \\Samuel~W.~ Remedios\inst{1} \and Fangxu Xing\inst{5} \and Jonghye Woo\inst{5} \and  Dzung~L.~Pham\inst{1,3} \and \\Aaron Carass\inst{1} \and Philip~V.~Bayly\inst{6} \and Jiachen Zhuo\inst{2} \and \\Ahmed Alshareef\inst{7} \and Jerry~L.~Prince\inst{1}}
\index{Bian, Zhangxing}
\index{Wei, Shuwen}
\index{Liang, Xiao}
\index{Lu, Yuan-Chiao}
\index{Remedios, Samuel W.}
\index{Xing, Fangxu}
\index{Woo, Jonghye}
\index{Pham, Dzung L.}
\index{Carass, Aaron}
\index{Bayly, Philip V.}
\index{Zhuo, Jiachen}
\index{Alshareef, Ahmed}
\index{Prince, Jerry L.}
\authorrunning{Bian et al.}
\institute{Johns Hopkins University, Baltimore MD, USA \\
    \and
    University of Maryland School of Medicine, Baltimore MD, USA \\ 
    \and
Uniformed Services University of the Health Sciences, Bethesda MD, USA \\
     \and
     The Henry M. Jackson Foundation for the Advancement of Military Medicine, Inc., Bethesda MD, USA \\
    \and
	Massachusetts General Hospital and Harvard Medical School, Boston MA, USA
    \and
    Washington University, St. Louis MO, USA \\ 
     \and
     University of South Carolina, Columbia SC, USA
}

\maketitle              % typeset the header of the contribution
\begin{abstract}
Magnetic resonance (MR) tagging is an imaging technique for noninvasively tracking tissue motion in vivo by creating a visible pattern of magnetization saturation~(tags) that deforms with the tissue.
Due to longitudinal relaxation and progression to steady-state, the tags and tissue brightnesses change over time, which makes tracking with optical flow methods error-prone.
Although Fourier methods can alleviate these problems, they are also sensitive to brightness changes as well as
spectral spreading due to motion.
To address these problems, we introduce the \underline{br}ightness-\underline{i}nvariant \underline{t}racking \underline{e}stimation~(BRITE) technique for tagged MRI.
BRITE disentangles the anatomy from the tag pattern in the observed tagged image sequence and simultaneously estimates the Lagrangian motion. The inherent ill-posedness of this problem is addressed by leveraging the expressive power of denoising diffusion probabilistic models to represent the probabilistic distribution of the underlying anatomy and the flexibility of physics-informed neural networks to estimate biologically-plausible motion. 
A set of tagged MR images of a gel phantom was acquired with various tag periods and imaging flip angles to demonstrate the impact of brightness variations and to validate our method.
The results show that BRITE achieves more accurate motion and strain estimates as compared to other state of the art methods, 
while also being resistant to tag fading. 

\end{abstract}
\section{Introduction}
Motion tracking using magnetic resonance imaging~(MRI) is used to understand tissue deformations and
organ biomechanics. 
Applications include analyzing cardiac motion~\cite{elsayedreview}, investigating musculoskeletal dynamics~\cite{moerman2012validation}, studying tongue movements during speech~\cite{yu2023spie, niitsu1994tongue, parthasarathy2007}, quantifying brain deformation during mild head impact~\cite{knutsen2014improved, bayly2021abe}, and evaluating liver strain~\cite{mannelli2012assessment}. 
Several imaging strategies including SPAMM~(Spatial Modulation of Magnetization) tagging~\cite{axel1989heart}, DENSE~(Displacement Encoding with Stimulated Echoes)~\cite{aletras1999dense}, and SENC~(Strain ENCoded) imaging~\cite{osman2001imaging} have been developed to aid this endeavor. 
Among these, SPAMM tagging has been widely adopted in clinical imaging due to
its low specific absorption rate and simplicity~\cite{elsayedreview} in application and interpretation.
SPAMM modulates the MR signal with periodic patterns, known as tags, that deform with the tissue.
This technique enables visual inspection and computational quantification of in vivo tissue deformation and strain.

\begin{figure}[!tb]
    \centering
    \includegraphics[width=1\linewidth]{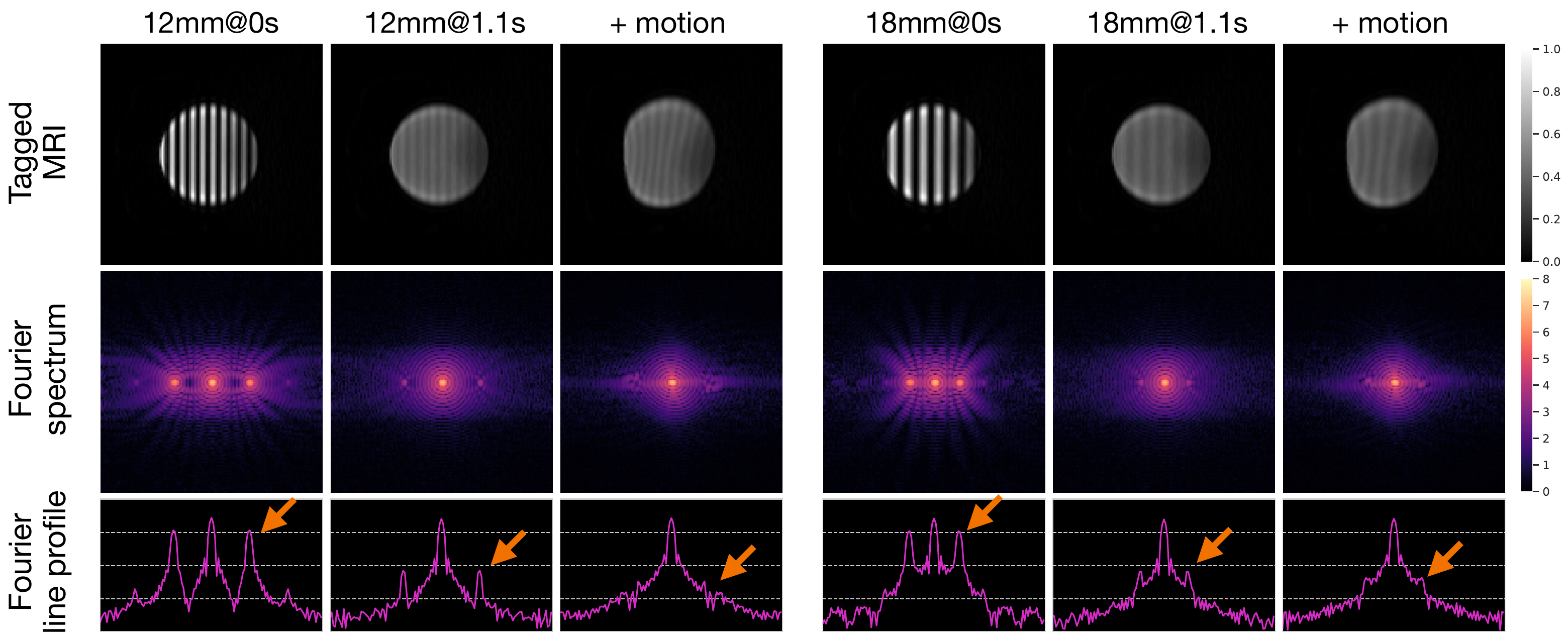}
    \caption{\textbf{Top row:} Tagged MRI. \textbf{Middle Row:} Their corresponding Fourier spectrum. \textbf{Bottom Row:} The midline profiles extracted from the Fourier spectra. The column names ``\textbf{Xmm@Ys}'' denote a tag period of X~mm acquired at Y~s following tagging.
    }
    \label{fig:intro_alias}
\end{figure}

\begin{figure}[!tb]
    \centering
    \includegraphics[width=1\linewidth]{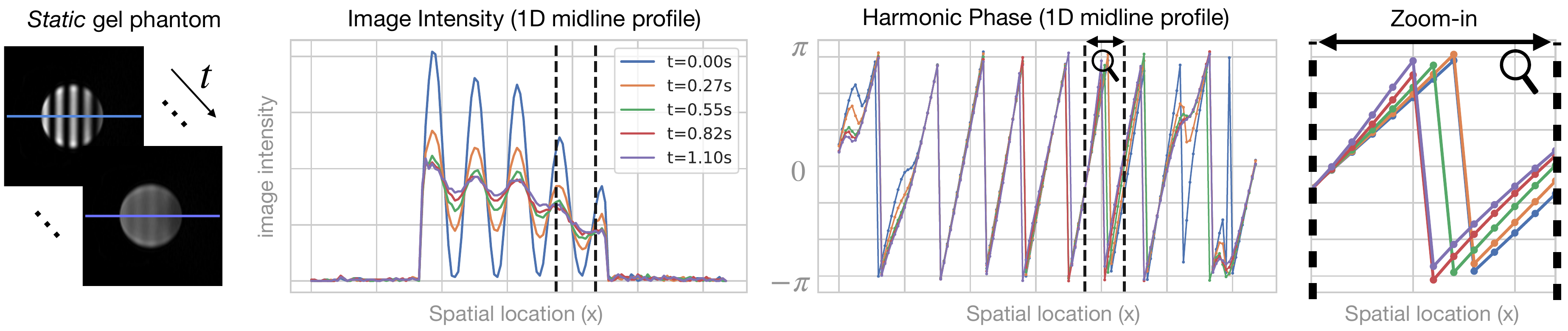}
    \caption{Tagged MRI of a \textit{static} phantom. Harmonic phases are distorted over time due to tag fading and spectral overlap, resulting in erroneous non-zero motion estimation.}
    \label{fig:distorted-harp}
\end{figure}

Methods for automated tissue-motion tracking of tagged MR images fall into two categories: (1)~Fourier-domain and (2)~image-domain analyses.
Among the Fourier-domain methods, HARP~(Harmonic Phase) analysis~\cite{osman1999cardiac} has emerged as a popular tag tracking method for SPAMM and related sequences~\cite{elsayedreview}.
The HARP technique and its descendants~\cite{PVIRA, mella2021harp, bian2024drimet} can
achieve sub-pixel accuracy.
Although image-domain methods using relaxed brightness constancy assumptions have been proposed~\cite{prince1992motion, gupta1995variable, dougherty1999validation}, recent work has shown that HARP is more accurate~\cite{elsayedreview, bian2024registering}. 
Despite HARP’s superiority, its accuracy remains limited because it depends on extracting the harmonic phase using a band-pass filter.
The resultant harmonic peaks are never perfectly isolated for the following reasons: (1)~Tag fading due to T1 relaxation and steady-state progression~\cite{bian2024registering} causes an increase in the central peak and a decrease in the harmonic peaks; (2)~Large tag periods bring the harmonic peaks closer together; and (3)~Non-rigid tissue motion deforms the tag lines, which broadens the harmonic peaks.
In addition, object rotation can cause the spectral peak to rotate entirely outside the band-pass region in Fourier space. Figure~\ref{fig:intro_alias} shows some of these effects on data from a silicone gel phantom. Figure~\ref{fig:distorted-harp} shows the impact of tag fading and spectral overlap on HARP.

The acquisition of additional tagged MR images has been used to avoid this problem.
The complementary SPAMM (CSPAMM)~\cite{CSPAMM1993} method suppresses the central peak by combining two SPAMM acquisitions; to yield both horizontal and vertical tags, four acquisitions are required.
The TruHARP~\cite{TruHARP} method uses five SPAMM acquisitions to produce completely isolated vertical and horizontal spectral peaks. 
Neither method is clinically feasible due to the additional data acquisition.  
Current practice in MR tagging is to use two SPAMM acquisitions, one horizontal and one vertical, or a single acquisition of grid tags.
In this paper, we address the spectral overlap problem in the presence of brightness changes when considering horizontal or vertical SPAMM tagged image sequences.  
 
To achieve better accuracy from tagged MRI in the presence of brightness changes, we propose the
\underline{br}ightness-\underline{i}nvariant \underline{t}racking \underline{e}stimation~(BRITE) method.  
The key innovation is to disentangle the anatomical signal (characterized by the central Fourier-domain peak) 
and the tag patterns (characterized by the harmonic peaks) given the observed tagged image sequence, 
while simultaneously estimating motion.
Our main contributions are as follows:
\textbf{(1)}~We analyze spectral overlap and tag fading in tagged MRI and quantify their impacts on motion tracking.
\textbf{(2)}~We introduce BRITE to disentangle anatomy from tags and account for tag fading while estimating Lagrangian motion.
\textbf{(3)}~Our physics-informed neural network~(PINN) is adapted for Lagrangian motion tracking with sequential images.
\textbf{(4)}~We validate BRITE using SPAMM-tagged MRI data of a silicone gel phantom acquired with various tag periods and imaging flip angles.

\section{Related Work}
\label{sec:related-work}

\parag{Fourier Domain approaches}
Fourier-based methods have long been used to extract motion information, typically via band-pass filtering.
As early as 1996, Zhang et al.~\cite{zhang1996fourier} extracted paired harmonic peaks to reconstruct and track horizontal and vertical tag lines. 
The HARP method~\cite{osman1999cardiac} tries to isolate a single harmonic peak with a band-pass filter. The inverse Fourier transform generates a phase image which allows for tracking of a tissue's phase---a material property of the tissue---over time.
Discontinuities in the phase image can cause interpolation issues which have been tackled with various approaches~\cite{PVIRA, mella2021harp, bian2024drimet}.
Qian et al.~\cite{qian2003segmenting, qian2006extraction} used convolution with Gabor filters to extract tag lines that can be tracked over time.
Arts et al.~\cite{arts2010mapping} proposed SinMod, which applies two skewed band-pass filters---one emphasizing low frequencies and the other high frequencies---to isolate a single harmonic peak.
From the resulting pair of complex images, the displacement is estimated as the phase difference between two time points divided by local frequency.
SinMod demonstrates good accuracy and robustness, especially at later time points where tag contrast diminishes.
All of these approaches rely on band-pass filters applied to inherently overlapping spectra, which can reduce motion-tracking accuracy.

\parag{Image domain approaches}
Prince and McVeigh~\cite{prince1992motion} proposed the variable brightness optical flow~(VBOF) method which relaxes the brightness constancy assumption by explicitly modeling tag fading through the imaging equation. 
The approach requires knowledge of the longitudinal relaxation time~(T1), transverse relaxation time~(T2), and spin density of the tissue.
Subsequent work~\cite{gupta1995variable} used a local linear intensity model, but imposed restrictions on the scaling field throughout the time course.
Dougherty et al.~\cite{dougherty1999validation} proposed an alternative method known as registration and change visualization~(RCV), which applied Laplacian ﬁltering to remove local offset and enhance edges. 
The underlying assumptions of RCV are viable only when a pair of images have minimal brightness variation.
Template-matching strategies have also been proposed~\cite{fisher1990automatic, amini1994mr} that have drawbacks due to large tag deformations~\cite{fisher1990automatic}, tag fading~\cite{fisher1990automatic, amini1994mr}, and require prior knowledge of the tissue and sequence parameters~\cite{amini1994mr}.

In the deep learning~(DL) era, Ye et al.~\cite{ye2021deeptag} introduced DeepTag, an unsupervised DL-based registration method that uses normalized cross-correlation as the similarity measure to directly register \textit{raw} tagged images without addressing tag fading.
While DeepTag can achieve rapid inference, it requires a sufficient number of tagged images from the target domain during training, thereby limiting its general applicability.
More recently, Bian et al.~\cite{bian2024registering} analyzed various commonly used image similarity measures for training DL-based registrations on \textit{raw} tagged images, and showed that tag fading remains a significant challenge.

Inspired by some of this previous work, BRITE disentangles the anatomy and tag patterns throughout the time course.
Unlike VBOF and template matching approaches, BRITE does not rely on prior knowledge of the imaging parameters; different from RCV, BRITE learns both the amplitude and offset of the tag pattern, and can handle large brightness variations; and unlike Deeptag, BRITE does not require tagged data to be collected for training.

\parag{A unified perspective}
Fourier- and image-domain approaches are fundamentally two sides of the same coin, related by the Fourier transform and its inverse.
Although spectral overlap effects are more directly observed in the Fourier domain, they ultimately influence the methods applied to either domain.

\section{Methodology}
\begin{figure}[!tb]
    \centering
    \includegraphics[width=1\linewidth]{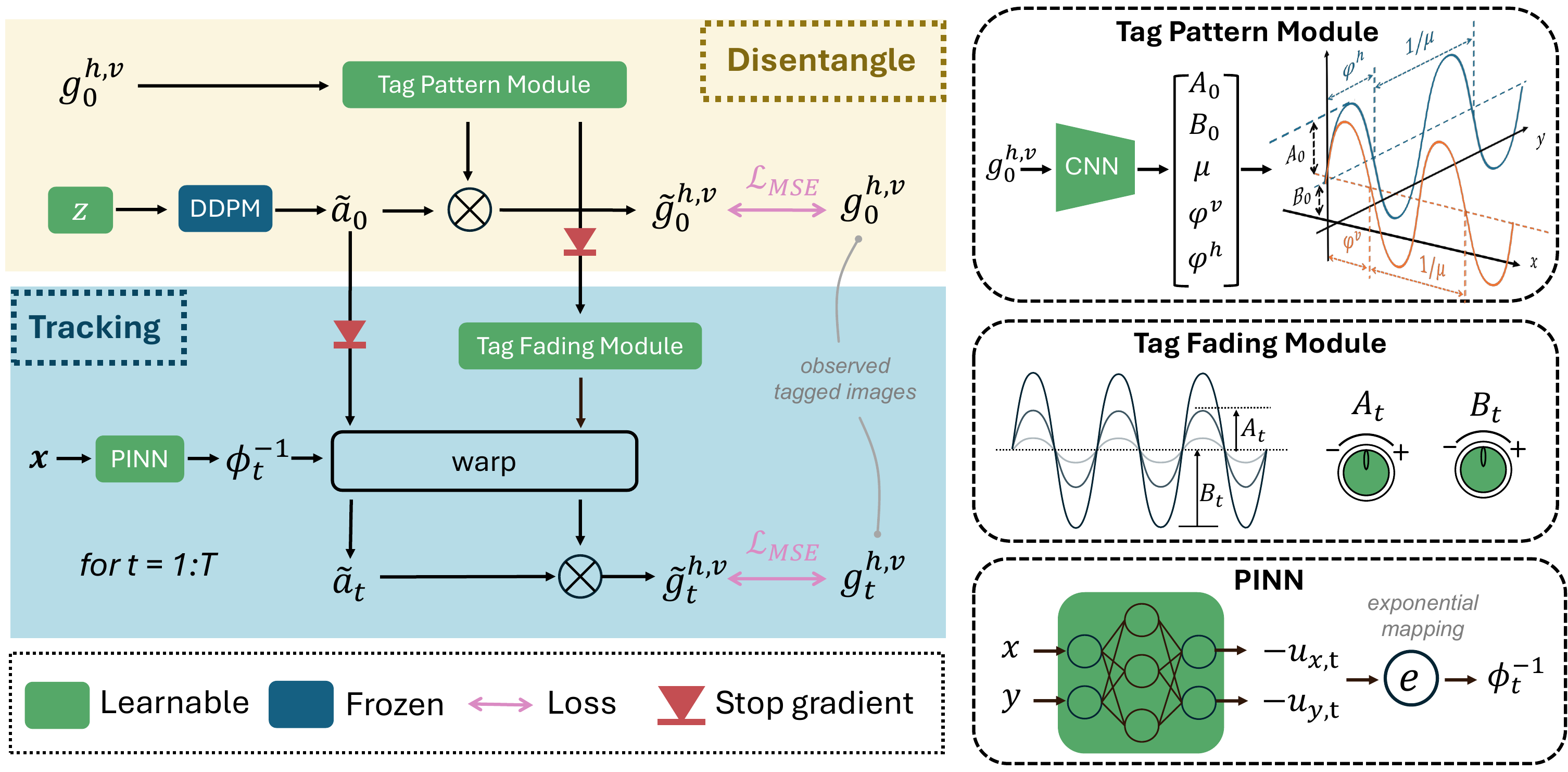}
    \caption{Overview of the Proposed BRITE Framework.
        The pipeline consists of two stages: disentangling and tracking. At $t = 0$, the ``Tag Pattern Module'' and a pretrained DDPM model are used to separate the anatomical image $\tilde{a}_{0}$ from the tag patterns $\tilde{p}^{h,v}_{0}$. Subsequently, for each time $t>0$, the PINN estimates the Lagrangian motion while the ``Tag Fading Module'' captures tag fading.}
    \label{fig:method}
\end{figure}
From a communication theory perspective, MR tagging can be partially viewed as the amplitude modulation of the anatomical signal with a \textit{low-frequency} carrier. This causes significant spectral overlap, making the reconstruction of the signal from the observed data an ill-posed problem. In our task, separating the harmonic and central spectral components (i.e., eliminating spectral overlap) is further complicated by tag fading and tissue motion.
In other words, from an image-domain perspective, we aim to simultaneously estimate the underlying anatomy, the fading tag pattern, and the tissue motion over time.

We first introduce the following notation. In 2D, $g$ denotes a ta\underline{g}ged image, $a$ an \underline{a}natomical image, and $p$ a tag \underline{p}attern. 
Superscripts $h,v$ indicate \underline{h}orizontal and \underline{v}ertical directions, respectively, and subscripts $t$ indicate time frame (e.g., $g_{t}^{v}$ denotes the vertically tagged image at time frame $t$).
When an equation applies to both directions, we write $\{h,v\}$ in the superscript for brevity.
The spatial variables $(x, y)$ are omitted whenever it does not cause ambiguity.

Figure~\ref{fig:method} shows an overview of the two key parts of BRITE, the disentanglement and the tracking.
Disentanglement estimates both the tag patterns and the underlying anatomical signal without tags. 
Tracking estimates tissue motion across time, incorporating the effects of tag fading at each frame.
Ultimately, this provides Lagrangian motion fields relative to the reference frame and yields separated anatomical and tag patterns that deform over time.

\parag{Disentangling Tagged Images} The forward model for MR tagging at $t = 0$ can be approximated by,
\begin{equation*} % write h and v equations compactly
    g_{0}^{\hvset} = a_{0} \otimes p_{0}^{\hvset},
\end{equation*}
where $g_{0}^{h}$ and $g_{0}^{v}$ are the observed tagged images with horizontal and vertical tags, respectively, $a_{0}(x,y)$ represents the unknown anatomical image at $t = 0$, and $\otimes$ denotes the element-wise product.

The functions $p_{0}^{h}(x, y)$ and $p_{0}^{v}(x, y)$ model the 1-1 SPAMM tag patterns,
\begin{equation*}
    p_{0}^{\hvset}(x,y\,; A_0, B_{0}, \mu, \varphi^{\hvset}) = A_{0} \sin(2\pi \mu x^{\delta^v} y^{\delta^h}  + \varphi^{\hvset}) + B_{0},
\end{equation*}
where $\varphi^h$, $\varphi^v$ denote the phases of the horizontal and vertical tag lines, respectively. 
$\mu$ is their spatial frequency, $A_{0}$ their initial amplitude, and $B_{0}$ their DC offset; these three parameters ($\mu$, $A_0$, and $B_0$) are shared between horizontal and vertical tags.
The indicator function is defined such that $\delta^{h}=1$ for the horizontal direction and $0$ otherwise; $\delta^{v}$ is defined similarly.
We use a \textit{randomly}-initialized CNN encoder $f_{\tagnet}$ parameterized by $\tagnet$, which predicts these parameters given the observed tagged images,
\begin{equation*}
    \{A_{0}, B_{0}, \mu, \varphi^{h} ,\varphi^v\} = f_{\tagnet} := f_{\tagnet}(\mathtt{cat}(g_{0}^{h}, g_{0}^{v})) \,,
\end{equation*}
where $\mathtt{cat}$ is channel-wise concatenation. 
Empirically, we found using this CNN-based prediction strategy helps optimization and convergence compared with directly optimizing parameters, which may be explained by deep image priors~\cite{ulyanov2018deep}.
We estimate the tag frequency $\mu$ rather than fixing it as a constant (e.g., read from the pulse sequence) to account for potential magnetic gradient imperfections.

To estimate the unknown anatomy $a_{0}$, we adopt a compressed sensing framework~\cite{bora2017compressed}.
We use a denoising diffusion probabilistic model~(DDPM) $\mathcal{G}(z)$, pretrained on anatomical MR images (e.g., cine or T1-weighted) to generate realistic candidate anatomies, which confines the search region for this ill-posed inverse problem. 
Solving 
\begin{equation}
    z^*, \tagnet^* = \argmin_{z, \tagnet} \sum_{i \in \{h,v\}} \left\| g_{0}^{i} - \mathcal{G}(z) \otimes p_{0}^{i}(;f_{\tagnet}) \right\|_2^2 \,
    \label{eq:opt-disentangle}
\end{equation}
yields the estimated anatomy $\tilde{a}_{0}$ and tag patterns $\tilde{p}_{0}^h, \tilde{p}_{0}^v$  that jointly explain the observed tagged images, where 
\begin{equation*}
    \tilde{a}_{0} = \mathcal{G}(z^*) \mbox{\quad and \quad}
    \tilde{p}_{0}^{\hvset} = p_0^{\hvset} \left( x,y\,; f_{\tagnet^*} \right) \,.
\end{equation*}

\parag{Tracking}
Having disentangled the initial frame, we now address subsequent time frames from $t=1$ to $t=T$.
Tag patterns fade and deform over time, and the anatomy should deform in the same way.
Our objective is to estimate tissue motion and evolving tag parameters for each $t$, ultimately providing a Lagrangian motion field referenced to $t = 0$.

We assume a deformation field $\phi_{t}$ that maps points from the reference frame ($t=0$) to their positions at time $t$.
We adapt a PINN~\cite{raissi2019physics} framework for tracking.
Specifically, a multilayer perception $f_{\pinnt}$ parameterized by \pinnt takes coordinates $(x,y)$ as input and outputs a stationary velocity field $\mathbf{u}_{t}(x,y)$, 
\begin{equation}
    \mathbf{u}_{t}(x,y) = f_{\pinnt}(x,y) \,.
    \label{eq:fpinn}
\end{equation}

We then integrate this velocity field and its negation via an fast exponential mapping~\cite{arsigny2006log}, also known as scaling and squaring, to obtain a biologically plausible diffeomorphic deformation and its inverse,
\begin{equation}
    \phi_{t}(x,y) = \exp(\mathbf{u}_{t})(x,y)\,, \quad \phi_{t}^{-1}(x,y) = \exp(-\mathbf{u}_{t})(x,y)\,.
    \label{eq:expu}
\end{equation}
The \textit{inverse} mapping allows us to warp the estimated anatomy to time $t$ as
\begin{equation}
    \tilde{a}_{t} = \tilde{a}_{0} \circ \phi_{t}^{-1}.
    \label{eq:tilde_a_t}
\end{equation}

To account for tag fading, we introduce a ``Tag Fading Module'' that has two learnable parameters $A_t$ and $B_t$ that model the time-varying amplitude and offset.
The faded tag patterns (before considering deformations) are,
\begin{equation*}
    \hat{p}_{t}^{\hvset}(x,y\,; A_t, B_t) = A_{t} \sin \left( 2\pi \tilde{\mu} x^{\delta^v} y^{\delta^h}  + \tilde{\varphi}^{\hvset} \right) + B_{t}\,.
\end{equation*}
Note that the tag frequency $\tilde{\mu}$ and phase $\tilde{\varphi}^{\hvset}$ are carried over from the solution to Equation~(\ref{eq:opt-disentangle}). 
Then incorporating the deformation gives the \textit{faded} and \textit{deformed} tag pattern at time $t$ as,
\begin{equation}
    \tilde{p}_{t}^{\hvset} = \hat{p}_{t}^{\hvset} \circ \phi_{t}^{-1} \,.
    \label{eq:tilde_p_t}
\end{equation}
Combining (4) and (5), the reconstructed tagged images at time $t$ are,
\begin{equation*}
    \tilde{g}_{t}^{\hvset} = \tilde{a}_{t} \otimes \tilde{p}_{t}^{\hvset} \,.
\end{equation*}
We jointly optimize \pinnt and $\{A_t,B_t\}$ to minimize the reconstruction error between $\tilde{g}_{t}^{h,v}$ and the observed $g_{t}^{h,v}$ at time $t$,
\begin{equation}
    \{A_{t}^*, B_{t}^*\}, \pinnt^* = \argmin_{\pinnt, \{A_{t}, B_{t}\}} \mkern9mu \sum_{i \in \{h,v\}} \left\| \tilde{g}_{t}^{i} - g_{t}^{i} \right\|_2^2 \,.
    \label{eq:opt-pinn}
\end{equation}
Once $\pinnt^*$ is obtained, the optimal Lagrangian motion field $\phi_t$ and its inverse $\phi_{t}^{-1}$ are computed from Equations~(\ref{eq:fpinn}) and~(\ref{eq:expu}).
To solve the motion at $t+1$, we initialize \pinntpo with $\pinnt^*$ and then optimize Equation~(\ref{eq:opt-pinn}).
This is an important feature of BRITE, as it prevents tag-jumping when the tissue’s Lagrangian motion exceeds half of the tag period.
Furthermore, unlike methods that compose motion between every pair of adjacent frames, BRITE avoids accumulating drifting errors.
Notably, no smoothness penalty is required for motion estimation. 

A brief delay (usually a few milliseconds) exists between tag preparation and the start of imaging, during which tissue may move and deform the tag lines. 
BRITE handles this by simply tracking from the first timeframe ($t=0$), so that the estimated $\phi^{-1}_0$ captures any pre-imaging deformation. Then for $t>0$, we substitute $\tilde{a}_0 \leftarrow \tilde{a}_0 \circ \phi^{-1}_0$ and $\hat{p}_{t}^{\{h,v\}} \leftarrow \hat{p}_{t}^{\{h,v\}} \circ \phi^{-1}_0$ into Equations~(\ref{eq:tilde_a_t}) and (\ref{eq:tilde_p_t}), and optimization of~(\ref{eq:opt-pinn}) proceeds as usual. We use this strategy for all experiments.

\parag{Implementation Details}
The function $f_{\tagnet}$ is a shallow 4-layer ResNet with basic residual blocks.
It accepts a two-channel input and outputs residual values for the tag-pattern parameters, which are added to user-specified initial values.
We initialize $A_0$ and $B_0$ at 0.45 and 0.55, respectively, both $\varphi^h$ and $\varphi^v$ at $2\pi$, and $\mu$ at the read from pulse setting.
The DDPM model uses a UNet backbone with three resolution levels (channels = (64, 64, 64)) and is pretrained on a synthetic dataset of random ovals (see Fig.~\ref{fig:method-proc}(b) for examples).
A DDIM~\cite{song2020denoising} scheduler with 20 steps is employed to sample the anatomy from the latent code $z$. 
The function $f_{\pinnt}$ is a fully connected network of three hidden layers, each with 128 neurons.
The number of steps for scaling and squaring is set to 7.
For the tag fading module, $(A_{t}, B_{t})$ are both initialized to 0.5.
The optimization of Equation~(\ref{eq:opt-disentangle}) runs for 600 iterations.
Equation~(\ref{eq:opt-pinn}) runs for 2,000 iterations per time frame.
All learnable parameters are trained with the Adam optimizer, using a learning rate of $1\times10^{-2}$ for $z$, $1\times10^{-4}$ for $f_{\tagnet}$ and $f_{\pinnt}$, and $5\times10^{-2}$ for $(A_{t}, B_{t})$. BRITE processes a 100-frame tagged MRI in 14 minutes~(8.4~s/frame), with a peak GPU memory usage of 5 GB on an Nvidia 4070 GPU.

\section{Experimental Setup}
\parag{Datasets}
We acquired tagged MRI data from a static cylindrical gel phantom~(made from Sylgard 527) using a 1-1 SPAMM tagging pulse followed by a Siemens FLASH imaging sequence. The total tagging angle was $90\degree$. The duration was 1.1~s with an 11~ms temporal resolution, yielding 100 time frames.
The imaging parameters were TR=3.67~ms, TE=1.63~ms, and a $128 \times 128 \times 1$ voxel matrix with a $2 \times 2 \times 10$~mm resolution.
Eight tagged sequences were acquired with tag periods of 9, 12, 18, and 26~mm and imaging flip angles of $5\degree$ and $10\degree$.
The choice of a $5\degree$ flip angle approximated the Ernst angle, assuming $T_1=900$~ms.
Simulated rigid rotation (N=1) and non-rigid deformations (N=20) were applied to each of the acquired tagged sequences (N=8) to provide ground-truth motion fields for evaluation. 
For non-rigid deformations, random displacements were generated at predefined control points and interpolated pixel-wise in between using B-splines. \looseness=-1

\parag{Evaluation Metrics}
We report the end-point error~(EPE), $\mathtt{EPE}(x,y) = \|\mathbf{d}_\mathrm{gt}(x,y) - \mathbf{d}_\mathrm{est}(x,y)\|_2$, the magnitude of the difference between the ground-truth and estimated displacement fields.
We also report the maximum principal strain~(MPS), which captures the largest principal strain experienced by the tissue. MPS reflects largest directional deformation which is of interest in biomechanical analyses.
The MPS error~(eMPS), defined as $\mathtt{eMPS}(x,y) = |\mathtt{MPS}_\mathrm{gt}(x,y) - \mathtt{MPS}_\mathrm{est}(x,y)|$, quantifies how close the ground truth and estimated strain fields are.
A Lagrangian reference frame was employed, with the first time frame being the reference.

\parag{Methods for Comparison}
We compared BRITE with five other methods.
We used the \methodit{HARP} software provided by the original authors.
The band-pass filter radius was set to half of the tagging frequency.
The source code of \methodit{SinMod} is not publicly available and thus we implemented it following the original publication.
The size of the squared cosine kernel used for smoothing was set to 15 and the exponent for quality measurement was set to 8.
For \methodit{SyN + raw} and \methodit{SyN + DRIMET}, we used the open-source {SyN}\footnote{\scriptsize \url{https://github.com/ANTsX/ANTs}} and the {DRIMET}\footnote{\scriptsize\url{https://github.com/jasonbian97/DRIMET-tagged-MRI}} implementation.
For both variants of \methodit{SyN}, we performed a grid search over different image similarity metrics ($\mathtt{CC}$ and $\mathtt{MeanSquares}$), various window sizes for $\mathtt{CC}$, and two motion-tracking strategies: (1)~registering adjacent frame pairs and composing transformations, and (2)~initializing the transformation with the previous time step's result and registering directly to the reference image. 
For \methodit{SyN + raw}, where raw tagged images are input, $\mathtt{CC}$ with a window size of 4 outperforms the default of 2.
For \methodit{SyN + DRIMET}, the $\mathtt{MeanSquares}$ metric yields the best results.
Both the raw and DRIMET inputs are multivariate: raw has two images (horizontal and vertical), while DRIMET has four (two per orientation), so we used SyN’s multivariate mode.
The composition strategy works best for both variants of \methodit{SyN}, and we report the best performance achieved under these parameter settings. 
We used the open-source algorithm \methodit{Deeptag}\footnote{\scriptsize\url{https://github.com/DeepTag/cardiac\_tagging\_motion\_estimation}}, which requires in-domain tagged data for training prior to testing.
We trained Deeptag on synthetic oval-shaped tagged images, which are subject to tag-fading and simulated deformations (following the same procedure as the test set).
Because Deeptag was originally designed for grid-tag patterns, we doubled the number of channels in the network’s first layer to accommodate both horizontal and vertical tagged images.

\section{Results and Discussion}
\begin{figure}[t!]
    \centering
    \includegraphics[width=0.9\linewidth]{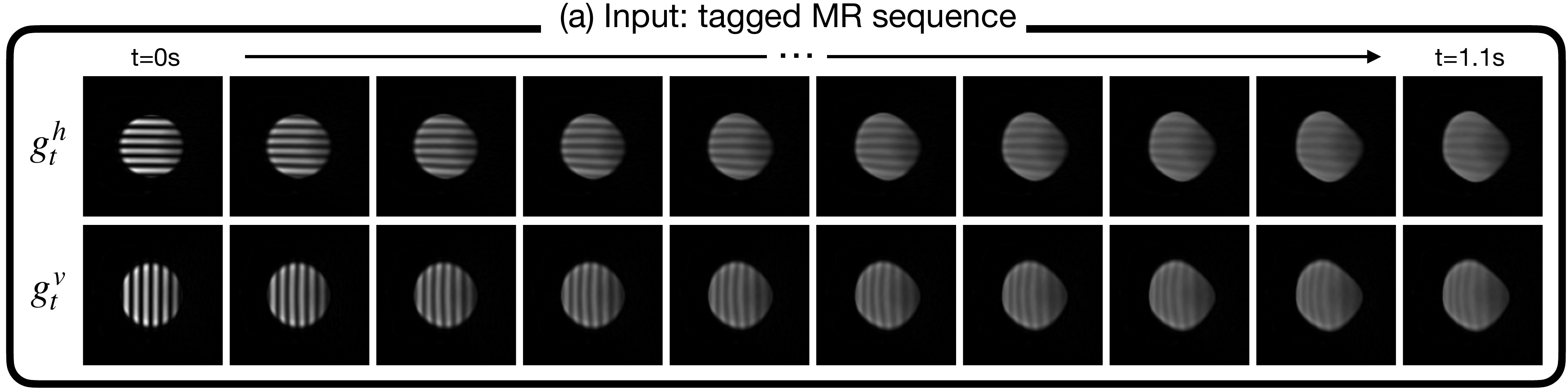}
    \includegraphics[width=0.9\linewidth]{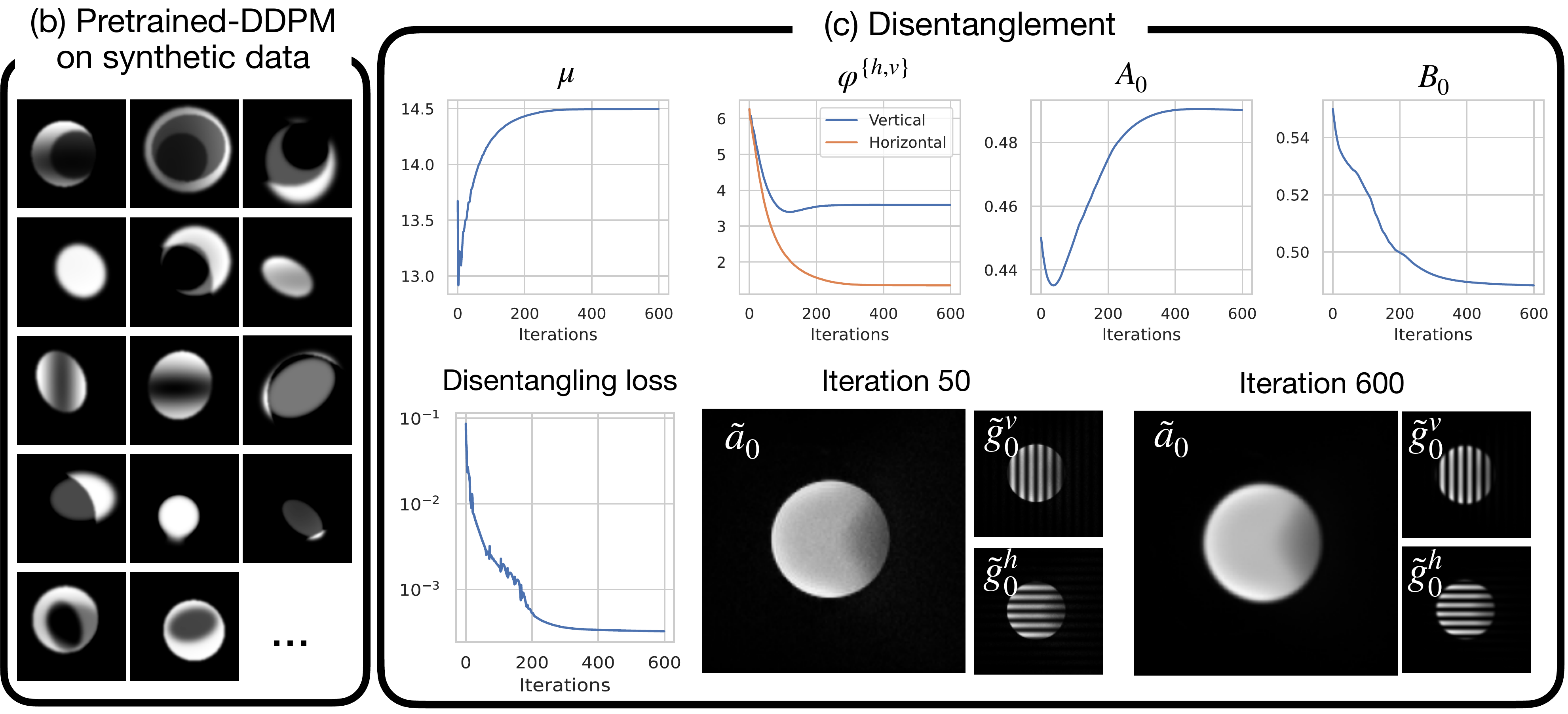}
    \includegraphics[width=0.9\linewidth]{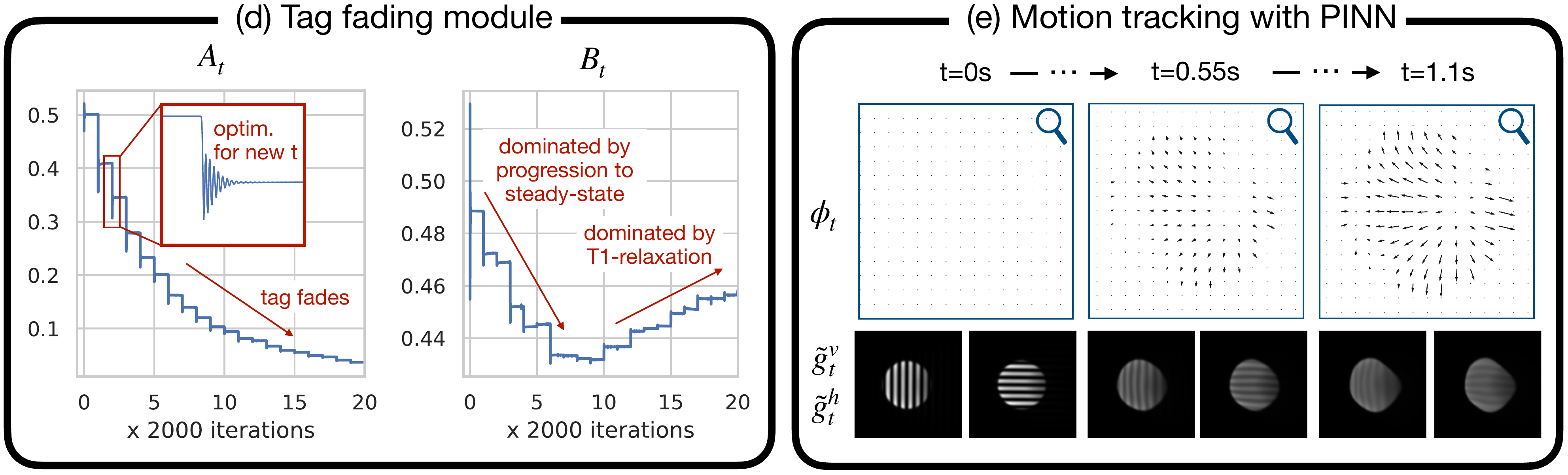}
    \caption{
    (a)~Input tagged MR sequence with horizontal and vertical tags.
    (b)~Synthetic oval-shaped images used for training the DDPM.
    (c)~Disentanglement process: shown are the optimization trajectories of $\mu, \varphi^h, \varphi^v, A_0, B_0$, the disentanglement loss, estimated anatomy $\tilde{a}_0$, and reconstructed tagged images $\tilde{g}_0$ at 50 and 600 iterations.
    (d)~Tag fading module optimization. Every 5th frame (total of 20 frames over 1.1s) is used for optimization and display clarity.
    (e)~Estimated Lagrangian motion fields. 
}
    \label{fig:method-proc}
\end{figure}
\parag{A closer look at BRITE. }
Figure~\ref{fig:method-proc} provides a detailed view of how each component of BRITE operates on a tagged MRI sequence undergoing tissue deformation and tag fading. 
Figure~\ref{fig:method-proc}(a)~shows the input sequence, where the contrast of the horizontal and vertical tags diminishes over time. 
Figure~\ref{fig:method-proc}(c)~illustrates the disentanglement step, which involves jointly optimizing the latent code $z$ and tag-pattern as per Equation~(\ref{eq:opt-disentangle}).
During optimization, the disentanglement loss decreases and tag pattern module converges.
Notably, even though the DDPM model is trained only on synthetic oval-shaped images (shown in Fig.~\ref{fig:method-proc}(b)), it successfully estimates the phantom’s underlying anatomy from the real tagged data.
As new time frames are introduced sequentially, the tag fading module and PINN work in tandem.
In~Fig.~\ref{fig:method-proc}(d), the tag fading module updates the amplitude and DC offset parameters ($A_{t}, B_{t}$), quickly adapting to changes in tag contrast caused by T1 relaxation and progression towards steady-state. 
Figure~\ref{fig:method-proc}(e) demonstrates motion estimation using the PINN, where the Lagrangian and diffeomorphic deformation are estimated throughout the sequence.

\begin{figure}[t!]
    \centering
    \includegraphics[width=0.95\linewidth]{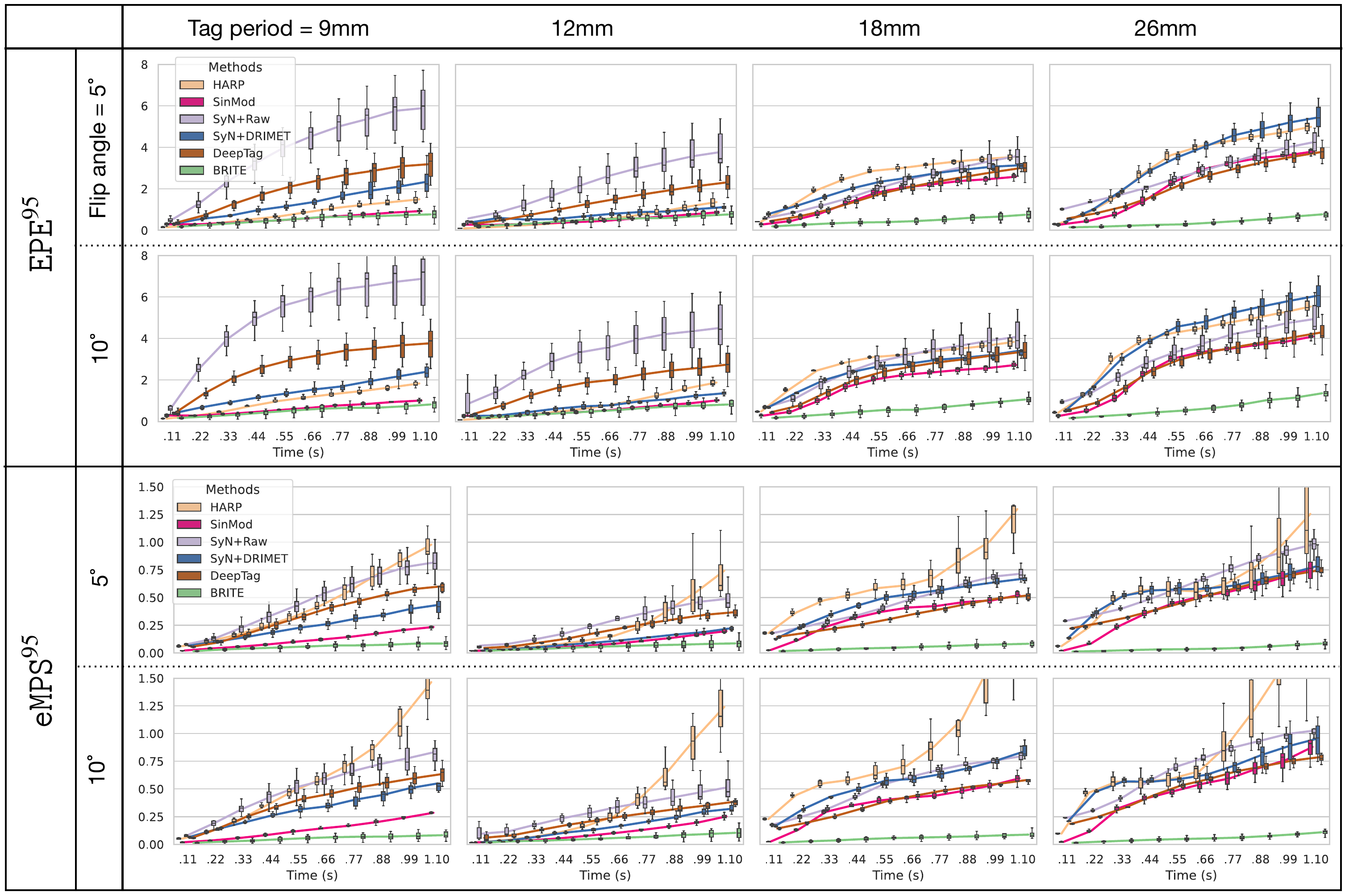}
    \caption{Non-rigid Deformation Evaluation. Results are shown for EPE (top two rows) and eMPS (bottom two rows) at different time points. Each column represents a different tag period, and each row corresponds to a flip angle.}
    \label{fig:nonrigid-boxplot}
\end{figure}

\begin{figure}[t!]
    \centering
    \includegraphics[width=0.9\linewidth]{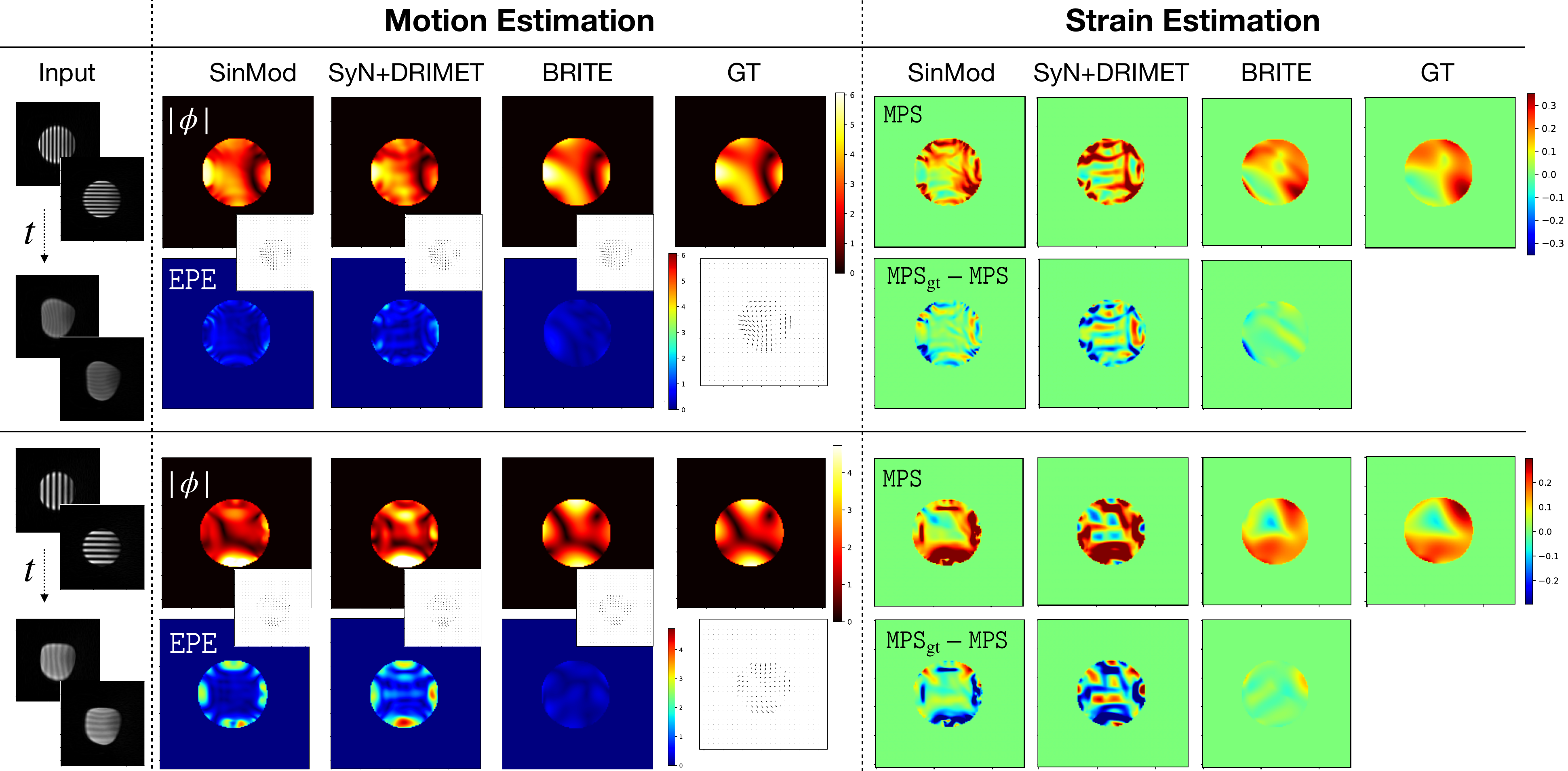}
    \caption{Qualitative Results under Non-Rigid Deformations. Two examples are shown with FA $= 5\degree$, TP $= 18$mm on the top and FA $= 10\degree$, TP $= 12$mm on the bottom. The left column displays the input images at $t = 0$s and $1.1$s. The Motion Estimation panel shows the displacement fields and its magnitude at $1.1$s, and EPE. The Strain Estimation panel shows the MPS fields and the corresponding errors. Only the well-performing methods are shown due to the limited space.}
    \label{fig:nonrigid-quali}
\end{figure}

\parag{Evaluation: Non-rigid Deformation.} We applied simulated non-rigid deformation fields to each tagged sequence acquired with varying imaging flip angles~(FA) and tag periods~(TP). 
Specifically, we considered FA $\in\{5^\circ, 10^\circ\}$ and TP $\in\{9, 12, 18, 26~\text{mm}\}$.
Figure~\ref{fig:nonrigid-boxplot} summarizes the performance of all methods. 
Figure~\ref{fig:nonrigid-quali} shows qualitative results.
\begin{figure}[tbhp]
    \centering
    \includegraphics[width=1.0\linewidth]{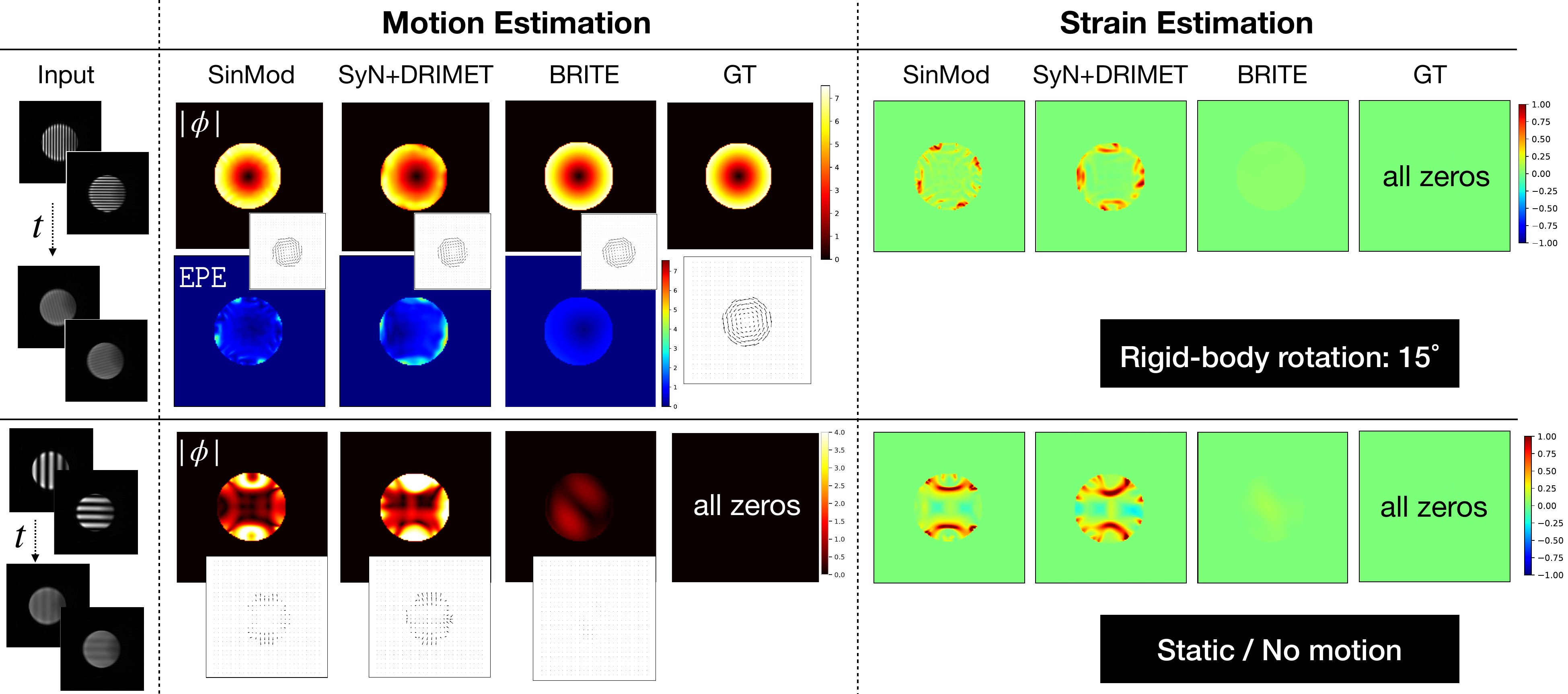}
    \caption{Qualitative Results under Rigid-Body Rotations and No Motion.
    The top row has FA $= 10\degree$, TP $ =9$mm for the $15\degree$ rigid-body rotation, and the bottom row shows FA $= 10\degree$, TP $= 26$mm for a static, no-motion scenario.}
    \label{fig:rigid-quali}
\end{figure}
BRITE generally outperforms all other methods in terms of \textit{both} motion~(EPE) and strain estimation accuracy~(eMPS).
SinMod generally ranks second but its accuracy degrades for larger tag periods, where spectral overlap becomes more pronounced.
Similar performance drops are observed in HARP-based approaches, including HARP itself and DRIMET, as they rely on Fourier-domain filtering, which is susceptible to spectral overlap as tags fade and the period increases.
Methods that operate directly on raw tagged images experience difficulties when tags fade, especially with smaller tag periods, where the denser tag pattern has a steeper sinusoidal profile.
A possible explanation is that, as tags fade, the algorithm must ``compress'' (or ``stretch'', depending on which image is fixed) the unfaded portion to maximize matching with the faded portion, creating erroneous motion.
A steeper sinusoidal profile exaggerates this error.
Comparing results across flip angles, we see that all methods generally perform better at $5^\circ$ than at $10^\circ$.
This likely reflects the fact that $5^\circ$ is closer to the Ernst angle, thus providing higher SNR and longer-lasting tag patterns for tracking.

The smoothness of deformation fields can significantly affect the MPS strain metric. For instance, HARP tracks each material point independently, making its strain measurements more noise-prone. In contrast, SinMod uses a squared cosine kernel for smoothing, while DRIMET, DeepTag, SyN, and BRITE apply regularization (e.g., smoothness terms or velocity-field integration) to produce smoother fields and less extreme strain values.

\parag{Evaluation: Rigid-body Rotation and No Motion}
All methods are evaluated on a $15\degree$ rigid-body rotation and static scenario.
BRITE generally performs best among all methods. 
Quantitative results are omitted due to limited space.
Qualitative examples are shown in Fig.~\ref{fig:rigid-quali}.

\parag{Summary} All methods show declining performance over time as the deformation grows and tags fade. Larger deformations broaden spectral peaks.
Fading tags reduce the tag contrast needed by intensity-based methods and exacerbate spectral overlap effects in Fourier-based methods.
BRITE suffers \textit{less} from these issues by disentangling the anatomy from the faded tags and applying its tag fading module independently at each time frame, ensuring a robust adaptation to variable brightness conditions.

\parag{Limitations \& Future Work}
Our current approach assumes a sinusoidal tag pattern (1-1 SPAMM or CSPAMM).
Future work will accommodate higher-order SPAMM and grid-tagging by modifying the tag model.
Although we only demonstrated our method’s performance in 2D, extending to 3D is straightforward.
Future work will validate BRITE on human organs.
\section{Conclusion}
We analyzed the phenomenon of tag fading and spectral overlap in tagged MRI and quantified their impacts on various motion tracking techniques. We introduced BRITE, a tracking approach that disentangles anatomy from tags and accounts for tag fading while estimating Lagrangian motion. We validated BRITE using SPAMM-tagged MR images of a silicone gel phantom acquired with various tag periods and imaging flip angles, demonstrating that BRITE is resistant to tag fading and provides more accurate motion and strain estimates compared to other state-of-the-art methods.

\textbf{Acknowledgment}
This study was partially supported by the National Institutes of Health~(NIH) grants U01NS112120, R01NS136056, R01DC018511, and National Science Foundation~(NSF) Graduate Research Fellowship (Grant No. DGE-1746891, Remedios). 

Under a license agreement between Myocardial Solutions and the Johns Hopkins University, Dr. Jerry L Prince and Johns Hopkins University are entitled to royalty distributions related to the HARP technology discussed in this publication. This arrangement has been reviewed and approved by the Johns Hopkins University in accordance with its conflict of interest policies. 

The opinions and assertions expressed herein are those of the authors and do not reflect the official policy or position of the Uniformed  Services University, the Henry M. Jackson Foundation for the Advancement of Military Medicine, or the Department of Defense. 

\bibliographystyle{splncs04}
\bibliography{main}
\end{document}